\documentclass[twocolumn,showpacs,showkeys,nofootinbib,aps,prd,letterpaper,floatfix,superscriptaddress]{revtex4-1}

\usepackage{amsmath,amssymb}
\usepackage{graphicx}
\usepackage{color}
\usepackage{url}

\def\nl{\nonumber\\}
\newcommand{\M}{\mathcal{M}}
\newcommand{\F}{\mathcal{F}}

\newcommand{\disc}{\textrm{disc}\,}
\newcommand{\BR}{\mathcal{B}\,}
\newcommand{\GeV}{{\,\textrm{GeV}}}
\newcommand{\beq}{\begin{equation}}
\newcommand{\eeq}{\end{equation}}
\newcommand{\mn}{M_{\pi^0}}
\newcommand{\mpn}{M_{\pi^0}^2}
\newcommand{\mpc}{M_\pi^2}
\newcommand{\mrr}{M_{\rho'}}
\newcommand{\diff}{\text{d}}
\newcommand{\J}{J/\psi}
\newcommand{\Jp}{J/\psi\to\pi^0}
\newcommand{\eps}{\epsilon}

\begin{document}

\title{Analysis of the \boldmath{$J/\psi\to\pi^0\gamma^*$} transition form factor}

\author{Bastian Kubis}\email{kubis@hiskp.uni-bonn.de}
\affiliation{Helmholtz-Institut f\"ur Strahlen- und Kernphysik (Theorie) and\\
             Bethe Center for Theoretical Physics,
             Universit\"at Bonn,
             53115 Bonn, Germany}

\author{Franz Niecknig}\email{niecknig@hiskp.uni-bonn.de}
\affiliation{Helmholtz-Institut f\"ur Strahlen- und Kernphysik (Theorie) and\\
             Bethe Center for Theoretical Physics,
             Universit\"at Bonn,
             53115 Bonn, Germany}

\begin{abstract}
In view of the first measurement of the branching fraction for $\Jp e^+e^-$ by the BESIII collaboration,
we analyze what can be learned on the corresponding transition form factor using dispersion theory.
We show that light-quark degrees of freedom dominate the spectral function, in particular two-pion 
intermediate states.  Estimating the effects of multipion states as well as charmonium, we arrive at a prediction
for the complete form factor that should be scrutinized experimentally in the future.
\end{abstract}

\pacs{11.55.Fv, 13.20.Gd, 13.75.Lb}

\keywords{Dispersion relations, Leptonic and radiative decays of the $\J$, Meson--meson interactions}

\maketitle

\section{Introduction}

The transition form factors of light vector mesons ($\omega$ and $\phi$ in particular) 
have garnered increased interest in the last few years due to their impact 
on the transition form factors of the lightest pseudoscalars~\cite{pi0TFF,etaTFF}, 
and hence on hadronic light-by-light scattering~\cite{roadmap}.
While e.g.\ the transition $\phi\to\eta e^+e^-$~\cite{Giovannella} seems compatible with 
a vector-meson dominance picture~\cite{Landsberg}, other experimental results, 
in particular for $\omega\to\pi^0\mu^+\mu^-$, seem to indicate strong deviations~\cite{NA60,NA60new},
which are hard to understand theoretically~\cite{omegapi-constraints}.

Recently the first measurements of the analogous transition form factors from charmonium into
light pseudoscalars have been reported by the BESIII collaboration~\cite{BES}, which has determined
the branching fractions for $\J\to P e^+e^-$, $P=\pi^0,\,\eta,\,\eta'$, and the transition form 
factor for the $\eta'$ final state.  The latter was parametrized in a simple monopole form~\cite{Fu}, 
with the scale found in the characteristic charmonium mass region.  
On the other hand, in Ref.~\cite{Fu}, such monopole form factors were assumed for all three final-state
pseudoscalars, and the corresponding branching fractions were estimated; interestingly enough, experiment 
agrees well with these predictions for $\eta$ and $\eta'$, while there seems to be a tension for
the $\pi^0$: the experimental determination arrives at 
$\BR(\Jp e^+e^-) = (7.56\pm1.32\pm0.50)\times 10^{-7}$~\cite{BES}, while
the theory prediction was $\BR(\Jp e^+e^-) = \big(3.89^{+0.37}_{-0.33}\big)\times 10^{-7}$~\cite{Fu}.

The assumption that the $q^2$-dependence of the $\Jp \gamma^*$ form factor should be determined by
the charmonium mass scale seems implausible, given that this would imply an isospin-breaking transition,
while the decay can proceed in an isospin-conserving manner, with the (virtual) photon being an isovector state,
hence dominated by light-quark degrees of freedom.  Indeed, it was pointed out by Chen \textit{et al.}~\cite{Chen} 
very recently, in an effective-Lagrangian-based analysis, that the contributions of light vector mesons
ought to be very sizable in this decay.  

In this article, we consider the $\Jp \gamma^*$ transition form factor, defined in Sec.~\ref{sec:def},
in dispersion theory.
Using the formalism employed previously for the analogous decays of the light isoscalar $\omega$ and
$\phi$ mesons~\cite{omegaTFF}, we show in Sec.~\ref{sec:disp} 
that it is dominated by the lightest, $\pi\pi$, intermediate
state, although not quite to the extent this dominance was found for $\omega$ and $\phi$.  
We give rough estimates of possible further light contributions beyond two pions, as well as from charmonium states.
While these induce a sizable uncertainty in the form factor, our results in Sec.~\ref{sec:results} show that
the experimentally observable decay spectra for $\Jp\ell^+\ell^-$, $\ell=e,\,\mu$, 
as well as the integrated branching fractions are rather stable, as they are dominated by the low-energy region.
We close with a summary.

\section{Definitions, kinematics}\label{sec:def}
The $\Jp \gamma^*$ transition form factor is defined according to
\begin{equation}\label{eq:fdef}
\langle \psi(p_V,\lambda)| j_\mu(0) | \pi^0(p) \rangle 
= - i \eps_{\mu\nu\alpha\beta}\eps^{\nu *}(p_V, \lambda) p^\alpha q^\beta f_{\psi\pi^0}(s),
\end{equation}
where $j_\mu$ denotes the electromagnetic current, $\lambda$ the polarization
of the $\J$ with $\eps^{\nu}(p_V, \lambda)$ the corresponding polarization vector, 
$q=p_V-p$, and $s=q^2$.  The form factor $f_{\psi\pi^0}(s)$ defined in this way
has dimension $\GeV^{-1}$.  Sometimes also the corresponding normalized 
form factor is used, denoted by
$F_{\psi\pi^0}(s) = {f_{\psi\pi^0}(s)}/{f_{\psi\pi^0}(0)} $.
The differential cross section for the decay $\Jp \ell^+\ell^-$, normalized
to the real-photon width, is given by
\beq \label{eq:dGamma}
\frac{\diff\Gamma_{\psi\to\pi^0\ell^+\ell^-}}{\Gamma_{\psi\to\pi^0\gamma}\,\diff s}
=\frac{16\alpha}{3\pi}\biggl(1+\frac{2m_\ell^2}{s}\biggr)
\frac{q_\ell(s)q_{\psi\pi^0}^3(s)}{(M_\psi^2-\mpn)^3}|F_{\psi\pi^0}(s)|^2,
\eeq
where $\alpha$ is the fine-structure constant, the real-photon width is determined by
\beq \label{eq:gammawidth}
\Gamma_{\psi\to\pi^0\gamma} = \frac{\alpha(M_\psi^2-\mpn)^3}{24 M_\psi^3}|f_{\psi\pi^0}(0)|^2,
\eeq
and the kinematical variables are 
\beq
q_\ell(s) = \frac{1}{2}\sqrt{s-4m_\ell^2}, \quad
q_{AB}(s) = \frac{\lambda^{1/2}(M_A^2,M_B^2,s)}{2\sqrt{s}},
\eeq
where $\lambda(a,b,c) = a^2+b^2+c^2-2(ab+ac+bc)$ is the usual K\"all\'en function.
The universal (QED) radiative corrections to~\eqref{eq:dGamma} have been calculated
in Ref.~\cite{radcorr}.

\section{Dispersive analysis}\label{sec:disp}

Dispersion theory attempts to reconstruct form factors from the corresponding discontinuity
across the cut along the positive real axis.  In principle, one would expect an 
unsubtracted dispersion relation to work for the $\Jp \gamma^*$ form factor, i.e.\
\beq
f_{\psi\pi^0}(s) = \frac{1}{2\pi i} \int_{4\mpc}^\infty \diff x \frac{\disc f_{\psi\pi^0}(x)}{x-s},
\eeq
where contributions to the discontinuity are given by multiparticle intermediate states
as well as single-particle pole contributions.
The lower limit of the integral is given by the lightest possible intermediate 
state, $\pi^+\pi^-$, that we will discuss in the following section.

\subsection{\boldmath{$\pi\pi$} intermediate states}

\begin{figure}
 \centering
\includegraphics[width= 0.6\linewidth,clip=true]{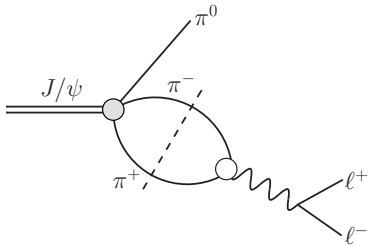} \\
\caption{Two-pion contribution to the discontinuity of the $\Jp\ell^+\ell^-$ 
transition form factor. 
The gray circle denotes the $\J\to3\pi$ $P$-wave amplitude, 
whereas the white circle represents the pion vector form factor.}
\label{fig:TFFdisc}
\end{figure}

The contribution of the two-pion intermediate state to the discontinuity of the $\Jp\gamma^*$ transition
form factor, see Fig.~\ref{fig:TFFdisc}, is given by~\cite{Koepp,omegaTFF}
\beq\label{eq:TFFdisc}
 \disc{f_{\psi\pi^0}^{\pi\pi}(s)}=\frac{i\,q_{\pi\pi}^3(s)}{6\pi \sqrt{s}}F_\pi^{V*}(s)f_1(s)\,\theta\big(s-4\mpc\big) ,
\eeq
where $F_\pi^V(s)$ is the pion vector form factor.
$f_1(s)$ is the projection of the $\J\to3\pi$ decay amplitude onto the $P$~partial wave:
with the amplitude $\M_{3\pi}=\M(\psi\to\pi^+(p_+)\pi^-(p_-)\pi^0(p_0))$ decomposed according to
\begin{equation}\label{eq:AmpMF}
 \M_{3\pi}=i\eps_{\mu\nu\alpha\beta}\eps^{*\mu} p_+^\nu p_-^\alpha p_0^\beta \F(s,t,u) ,
\end{equation}
it is given by
\begin{equation}
 f_1(s)=\frac{3}{4}\int_{-1}^{1} \diff z \big(1-z^2\big)\, \F(s,t,u) , \label{eq:f1-angint}
\end{equation}
where $z=(t-u)/(4q_{\pi\pi}(s)q_{\psi\pi^0}(s))$, and $s=(p_++p_-)^2$, $t=(p_-+p_0)^2$, $u=(p_++p_0)^2$.

To describe the $\J\to3\pi$ amplitude, we rely on the phenomenological
observation that the Dalitz plot for this decay is entirely dominated by $\pi\rho$ intermediate
states, i.e.\ by the lowest resonance in the $\pi\pi$ $P$~wave; neither higher resonances, nor
resonant higher partial waves are observed~\cite{BES:JPsi3pi}.  We do not attempt to explain
this suppression of additional structures~\cite{Szczepaniak:Veneziano}, but just take it as the 
starting point for a generalized partial-wave decomposition that stops at $P$-wave contributions 
only~\cite{Guo:Jpsi,V3pi},
\begin{equation}\label{eq:reconstruction}
 \F(s,t,u)=\F(s)+\F(t)+\F(u) .
\end{equation}
Final-state interactions between all three pions are implemented in a Khuri--Treiman-type 
formalism~\cite{KhuriTreiman}, leading to~\cite{V3pi} (compare also Ref.~\cite{Danilkin})
\begin{align}\label{eq:inteeqn}
\F(s)&=a\, \Omega(s)\biggr\{1 + \frac{s}{\pi}\int_{4\mpc}^{\infty}\frac{\diff x}{x}\frac{\sin\delta(x)\hat\F(x)}{|\Omega(x)|(x-s)}\biggr\} ,\nl
 \hat\F(s)&=3\langle(1-z^2)\F\rangle(s) ,
\end{align}
where $\delta(s)\doteq \delta_1^1(s)$ is the isospin 1 $\pi\pi$ $P$-wave phase shift
taken from Refs.~\cite{Madrid,CCL} and
$\langle . \rangle$ denotes angular averaging. $\Omega(s)$ is the Omn\`es function calculated
from the phase shift $\delta(s)$,
\beq\label{eq:Omnes}
\Omega(s) = \exp\bigg\{ \frac{s}{\pi}\int_{4\mpc}^\infty \diff x\frac{\delta(x)}{x(x-s)}\bigg\}.
\eeq
The function $\hat\F$ denotes the partial-wave 
projection of the crossed-channel contributions, which are fed into the dispersive solution for $\F$.
The partial wave $f_1(s)$ is related to both by $f_1(s)=\F(s)+\hat\F(s)$.
The single subtraction constant $a$ only affects the overall normalization of the amplitude and can be 
fixed, up to a phase, from the total $\J\to3\pi$ branching fraction.  
For the pion vector form factor $F_\pi^V(s)$, we also employ a representation based on the Omn\`es 
function~\eqref{eq:Omnes}; see Ref.~\cite{omegaTFF} for details.
This fully determines~\eqref{eq:TFFdisc}.

In particular, we can calculate the two-pion contribution to the \emph{real}-photon transition 
$\Jp \gamma$ in the form of a sum rule~\cite{omegaTFF}:
\beq \label{eq:TFFsumrule}
f_{\psi\pi^0}^{\pi\pi}(0)  
=\frac{1}{12\pi^2}\int_{4\mpc}^{\infty}\diff x\frac{q_{\pi\pi}^3(x)}{x^{3/2}}F_\pi^{V*}(x)f_1(x) .
\eeq
As the partial wave $f_1(s)$ depends on an unknown overall normalization constant $a$, the cleanest
prediction following from~\eqref{eq:TFFsumrule} is in principle 
the ratio $\BR(\Jp\gamma)/\BR(\J\to3\pi)$~\cite{omegaTFF}, which is determined by
the phase shift $\delta(s)$ alone.

The experimental branching fraction for $\Jp\gamma$~\cite{PDG}, together with~\eqref{eq:gammawidth}, leads to
$|f_{\psi\pi^0}(0)|= (6.0 \pm 0.3)\times 10^{-4}\GeV^{-1}$, whereas the sum rule~\eqref{eq:TFFsumrule}
results in
\beq \label{eq:N2pi}
|f_{\psi\pi^0}^{\pi\pi}(0)| = (4.8 \pm 0.2)\times 10^{-4}\GeV^{-1},
\eeq
where the uncertainty is a combination of the experimental uncertainties in $\BR(\J\to3\pi)$
and the width of the $\J$,  
as well as the one in the dispersive integral.
We therefore conclude that the two-pion intermediate state alone saturates the sum rule for 
the transition form factor normalization to about 80\%.
Note that this presents a very significant enhancement over a simple vector-meson dominance estimate
based on the decay chain $\J\to \rho^0\pi^0$, $\rho^0\to\gamma$ (see e.g.\ Ref.~\cite{Chen}), which would rather result
in $|f_{\psi\pi^0}^{\rho}(0)| \approx 3.3\times 10^{-4}\GeV^{-1}$.

This result is to be compared to similar sum rules for the decays $\omega\to\pi^0\gamma$ and
$\phi\to\pi^0\gamma$, which were observed to be saturated to more than 90\% accuracy~\cite{omegaTFF}.
The difference looks rather plausible, as the branching fractions of the $\J$ 
into more than three pions are actually \emph{larger}:
$\BR(\J\to3\pi) = (2.11\pm0.07)\%$, 
$\BR(\J\to5\pi) = (4.1\pm0.5)\%$, 
$\BR(\J\to7\pi) = (2.9\pm0.6)\%$~\cite{PDG}. 
It would therefore not come as a surprise if the inelastic contributions to the $\Jp\gamma^*$ 
transition form factor, coming from the discontinuities due to four and more pions,
played a much more significant role than e.g.\ for the $\omega\to\pi^0\gamma^*$
transition.  However, the information on the branching fractions alone does not lend itself easily
to an improvement of the radiative decay/the transition form factor before more differential information
in the form of a partial-wave analysis becomes available.
From data on $e^+e^-\to [\text{hadrons}]_{I=1}$, the most important inelastic intermediate state 
of isospin $I=1$ ought to 
be $4\pi$, which couples strongly to the $\rho'(1450)$ resonance. In a very simplistic model approach,
we therefore add a $\rho'(1450)$ resonance to the $\Jp\gamma^*$ transition form factor as an approximation
to the possible effects of multipion intermediate states, which we allow to contribute
between 10\% and 30\% of the dominant $\pi\pi$ channel to the sum rule for $f_{\psi\pi^0}(0)$.
Note that in a more complete/realistic description the dispersive contributions from $\pi\pi$ and 
inelastic states would have to be treated as coupled channels; see e.g.\ Ref.~\cite{pionvff_Hanhart} for a
corresponding analysis of the pion vector form factor.  
We reconstruct the $\rho'(1450)$ propagator dispersively from the imaginary part of an energy-dependent
Breit--Wigner function,
\begin{align} \label{eq:rhopBW}
\disc f_{\psi\pi^0}^{\rho'}(s) &= \frac{2i\sqrt{s}\mrr^2\Gamma_{\rho'}(s)}{(\mrr^2-s)^2+s\,\Gamma_{\rho'}^2(s)} , \nl
\Gamma_{\rho'}(s) &= \bigg(\frac{\mrr^2}{s}\bigg)^2 \,\bigg[\frac{s-16\mpc}{\mrr^2-16\mpc}\bigg]^{7/2}\Gamma_{\rho'}\big(\mrr^2\big)  \nl
& \qquad \times \theta\big(s-16\mpc\big) ,
\end{align}
thus maintaining a reasonable analytic behavior.  $\Gamma_{\rho'}(s)$ reproduces the near-threshold behavior
of four-pion phase space~\cite{pionvff_Hanhart}. 
With $\mrr = 1.6\GeV$, $\Gamma(\mrr^2) = 0.6\GeV$, the dispersive integral over~\eqref{eq:rhopBW} 
results in a function of which the peak
position and width agree with the Particle Data Group Breit--Wigner parameters~\cite{PDG}.

\subsection{Light isoscalar contributions to \boldmath{$\J\to\eta,\eta'\gamma^*$}}

Given the strong impact of light-quark degrees of freedom on the $\Jp\gamma^*$ transition, to what 
extent may something similar be true for the decays $\J\to\eta,\eta'\gamma^*$?  In the limit of isospin
conservation, here only the light \emph{isoscalar} vector mesons $\omega$ and $\phi$ can contribute, 
which in the context of this study we consider as sufficiently narrow that we can approximate their contribution
to the discontinuity by $\delta$-functions,
\beq
\disc f_{\psi P}^V(s) = 2\pi i \sum_{V=\omega,\phi} c_{PV} F_V M_V \delta(s-M_V^2) ,
\eeq
where $P=\eta,\,\eta'$.  Here, $F_V$ denote the vector-meson decay constants, determined from the corresponding
electron--positron decay rates by
\beq \label{eq:ee}
\Gamma_{V\to e^+e^-} = \frac{4\pi\alpha^2}{3} \frac{F_V^2}{M_V} 
\eeq
(neglecting the mass of the electron), while the effective coupling constants $c_{PV}$
are fixed from the decay rates $\J\to PV$ by
\beq \label{eq:JPV}
\Gamma_{\J\to PV} = \frac{|c_{PV}|^2}{96\pi M_\psi^3} \lambda^{3/2}\big(M_\psi^2,M_{V}^2,M_P^2\big) .
\eeq
We do not attempt a symmetry-based analysis of the couplings $c_{PV}$ here (compare Refs.~\cite{Chen,Escribano,Zhao} 
and references therein), but just estimate them individually from data; we note that SU(3) symmetry suggests
constructive interference of $\omega$ and $\phi$ contributions for the $\eta$ final state, but 
destructive interference for the $\eta'$. 
Individually, the estimated contributions of $\omega$ and $\phi$
to the transition form factors at the real-photon point, given simply by $f_{\psi P}^V(0) = c_{PV}F_V/M_V$, amount to
\begin{align}
\big|f_{\psi\eta}^{\{\omega,\phi\}}(0)\big| &\approx \{0.9,\,0.8\}\times 10^{-4}\GeV^{-1} , \nl
\big|f_{\psi\eta'}^{\{\omega,\phi\}}(0)\big| &\approx \{0.3,\,0.7\}\times 10^{-4}\GeV^{-1} , 
\end{align}
whereas the decay rates for $\J\to\eta,\eta'\gamma$~\cite{PDG} suggest
$|f_{\psi\eta}(0)| = (35\pm1)\times 10^{-4}\GeV^{-1}$,
$|f_{\psi\eta'}(0)| = (85\pm3)\times 10^{-4}\GeV^{-1}$.
We conclude, in accordance with Ref.~\cite{Chen}, 
that for the isoscalar transition form factors, light-quark resonances
contribute only at the percent level, so the corresponding spectral functions are entirely
dominated by charmonium intermediate states, in the loose sense of both $c\bar{c}$ resonances and
open charm--anticharm continuum contributions.

\subsection{Estimate of charmonium contributions}

Given the vast dominance of charmonium in the transition form factors for $\eta$ and $\eta'$,
we may wonder if such effects cannot also be sizable for $\Jp\gamma^*$, even though in that case,
they break isospin symmetry.  Indeed, in the same narrow-width approximation employed in the previous
section, we can determine the contribution specifically of the $\psi(2S)$, using experimental information
on the branching fractions for $\psi(2S)\to \J\pi^0$ and $\psi(2S)\to e^+e^-$~\cite{PDG} and analogous relations
to~\eqref{eq:ee} and \eqref{eq:JPV} to determine the decay constant $F_{\psi(2S)}$ and
an effective coupling $c_{\pi^0\psi(2S)}$.  Surprisingly, one finds
\beq
|f_{\psi\pi^0}^{\psi(2S)}(0)| = (5.3 \pm 0.1)\times 10^{-4}\GeV^{-1},
\eeq
which is larger than the two-pion contribution~\eqref{eq:N2pi}.
However, the comparison to the $\J\to\eta\gamma^*$ transition form factor
demonstrates that this observation is too simplistic.  Here, branching fractions into $\J\eta$ 
(and $e^+e^-$) are known for the excited charmonium resonances $\psi(2S)$, $\psi(3770)$, and 
$\psi(4040)$, so we can determine their contributions to 
the sum rule for $f_{\psi\eta}(0)$.  Their moduli turn out to be $(117\pm2)\times 10^{-4}\GeV^{-1}$, 
$(25\pm6)\times 10^{-4}\GeV^{-1}$, and $(70\pm7)\times 10^{-4}\GeV^{-1}$, respectively, compared
to the total $|f_{\psi\eta}(0)| = (35\pm1)\times 10^{-4}\GeV^{-1}$.  We conclude
that there need to be strong cancellation effects between different charmonium resonances (as well
as, probably, open-charm continuum channels) in the $\J\to\eta\gamma^*$ form factor spectral function
in order to explain the observed rate for $\J\to\eta\gamma$.  

To estimate the total charmonium contribution to $\Jp\gamma$, $|f_{\psi\pi^0}^{c\bar c}(s)|$, 
we therefore assume that the ratio of $\psi(2S)$ contributions to the transitions into $\pi^0$ and $\eta$ 
gives a useful indication of the ratio of overall $c\bar{c}$ effects:
\beq
0.01 \lesssim
\frac{|f_{\psi\pi^0}^{c\bar c}(0)|}{|f_{\psi\eta}^{c\bar c}(0)|} \lesssim 
\frac{|f_{\psi\pi^0}^{\psi(2S)}(0)|}{|f_{\psi\eta}^{\psi(2S)}(0)|} \approx 0.045 .
\eeq
We assume this to be an upper limit due to the observation that the $\psi(2S)\to\J\pi^0$ decay
rate is somewhat enhanced relative to $\psi(2S)\to\J\eta$ due to charmed-meson loop 
effects~\cite{Guo:JpsiQuarkMasses}.  
The lower limit of 1\% is the size of a typical, nonenhanced isospin-breaking effect,
which requires cancellation of individual charmonium resonances by no more than one order of magnitude.
We therefore estimate (with $|f_{\psi\eta}^{c\bar c}(0)| \approx |f_{\psi\eta}(0)|$)
\beq
0.3 \times 10^{-4}\GeV^{-1} \lesssim |f_{\psi\pi^0}^{c\bar c}(0)| \lesssim 1.6\times 10^{-4}\GeV^{-1}.
\eeq
For the $s$-dependence of this contribution, we adopt the simple monopole ansatz~\cite{Fu},
\beq \label{eq:monopole}
f_{\psi\pi^0}^{c\bar c}(s) = \frac{f_{\psi\pi^0}^{c\bar c}(0)}{1-s/\Lambda^2},
\eeq
and vary the effective scale $\Lambda$ between the mass of the $\J$ and the mass of the $\psi(2S)$.

\section{Results and discussion}\label{sec:results}

\begin{figure}
\includegraphics[width=\linewidth]{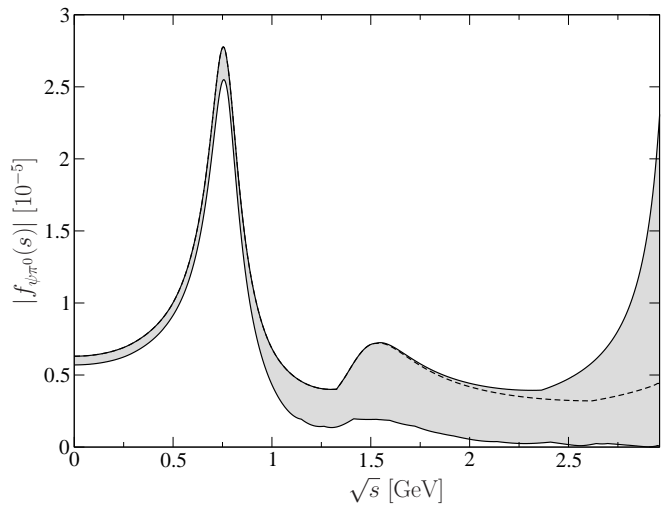}
\caption{Modulus of the transition form factor $|f_{\psi\pi^0}(s)|$.
See the main text for the discussion of the uncertainty band.  
The dashed curve denotes the upper limit of the band in the case that the 
scale $\Lambda$ for the charmonium contribution is fixed to the $\psi(2S)$ mass.}
\label{fig:TFF}
\end{figure}

In Fig.~\ref{fig:TFF}, we show the modulus of our total form factor
\beq
f_{\psi\pi^0}(s) = f_{\psi\pi^0}^{\pi\pi}(s)+f_{\psi\pi^0}^{\rho'}(s)+f_{\psi\pi^0}^{c\bar c}(s).
\eeq
While $f_{\psi\pi^0}^{\pi\pi}(s)$ is fixed within its (rather narrow) uncertainty, we vary
the effective $\rho'$ and charmonium contributions within the rather generous error bands discussed
in the previous sections, with unknown relative signs, but 
subject to the constraint that the $\Jp\gamma$ sum rule be fulfilled within
experimental uncertainties, $|f_{\psi\pi^0}(0)| = (6.0\pm0.3)\times10^{-4}\GeV^{-1}$.
This variation in the normalization determines the error band in the form factor at low energies, 
while the theoretical variation within our rather crude estimates of the $\rho'$ and $c\bar c$ 
contributions dominates the uncertainty above $\sqrt{s} \gtrsim 1\GeV$.
While all the light-quark resonance contributions drop like $1/s$ above their respective characteristic
scales (the masses of $\rho$ and $\rho'$), $f_{\psi\pi^0}^{c\bar c}(s)$ rises close to the upper limit of 
the accessible decay phase space and dominates the total form factor there. 
In particular, if the characteristic scale $\Lambda$ is set to the $\J$ mass, 
$f_{\psi\pi^0}^{c\bar c}(s)$ is enhanced by roughly a factor $M_\psi/(2\mn)\approx 11.5$ at $\sqrt{s}=M_\psi-\mn$.
Figure~\ref{fig:TFF} also shows the upper form factor limit using $\Lambda = M_{\psi(2S)}$ only, which
limits the rise significantly.

\begin{figure*}
\includegraphics[height=6.5cm, clip=true]{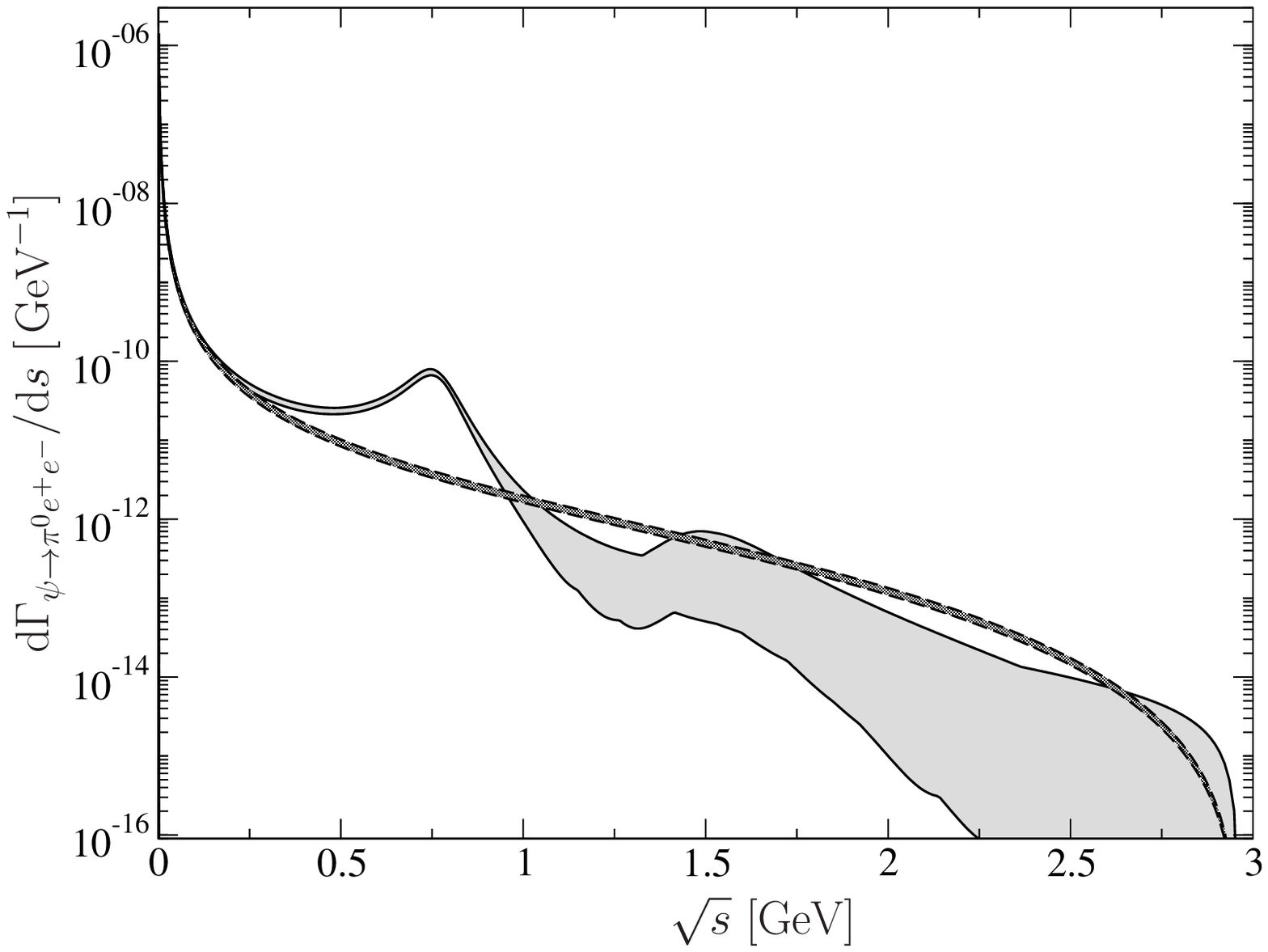} \hfill 
\includegraphics[height=6.5cm, clip=true]{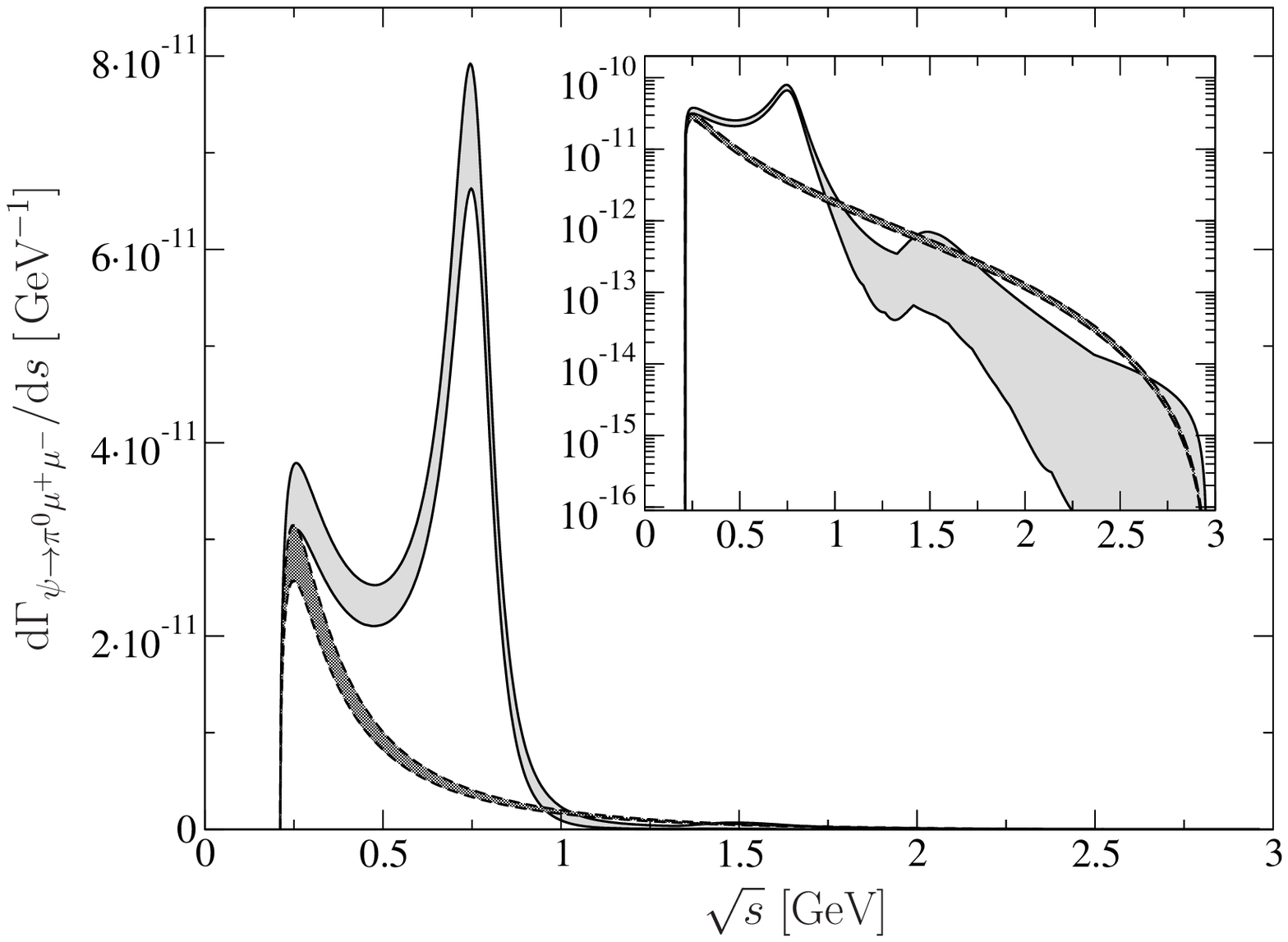} 
\caption{Differential decay rates $\diff\Gamma/\diff s$ for $\Jp e^+e^-$ (left) and $\Jp \mu^+\mu^-$ (right);
the insert in the right panel displays the same distribution on a logarithmic scale.  The full gray bands 
correspond to our form factor prediction, while the dashed bands show the QED distributions for comparison, 
i.e.\ with the form factor set to a constant.}
\label{fig:dG}
\end{figure*}
In addition, Fig.~\ref{fig:dG} shows the resulting differential decay rates for 
$\Jp e^+e^-$ and $\Jp \mu^+\mu^-$.  For comparison, we also display the distributions
obtained by setting $f_{\psi\pi^0}(s) \equiv f_{\psi\pi^0}(0)$. 
For both final states, the clear enhancement due to the $\rho$ resonance in the $\pi\pi$ intermediate
state is the dominating form factor feature, while $\diff\Gamma_{\psi\to\pi^0e^+e^-}/\diff s$ rises strongly
near $\sqrt{s}=0$.  The sizable form factor uncertainty at large energies occurs in a region where 
phase space already suppresses the decay distributions strongly; in particular, a potential strong rise 
in the form factor due to $\J$ pole contributions is probably not experimentally observable.

Integrating over the respective spectra, we can calculate the branching fractions for the two dilepton
final states.  We find
\begin{align}\label{eq:BR}
\BR(\Jp e^+e^-) &= (5.5 \ldots 6.4)\times 10^{-7} ,\nl
\BR(\Jp \mu^+\mu^-) &= (2.7\ldots 3.3)\times 10^{-7}.
\end{align}
This can be compared to the numbers obtained from QED spectra with a constant form factor,
$\BR(\Jp e^+e^-) = (3.7\pm0.4)\times 10^{-7}$,
$\BR(\Jp e^+e^-) = (0.9\pm0.1)\times 10^{-7}$.
A monopole form factor as in~\eqref{eq:monopole}, with the scale given by the mass of the $\psi(2S)$~\cite{Fu}, 
magnifies these by a few percent only.
Our dispersive result therefore enhances the branching fractions very considerably, almost by a factor of
3 for the muon final state.
Note that the dispersive prediction~\eqref{eq:BR} is remarkably stable due to the dominance of the
low-energy region in the integrated rate.

It is rather unclear how to compare~\eqref{eq:BR} to the experimental result 
$\BR(\Jp e^+e^-) = (7.56\pm1.32\pm0.50)\times 10^{-7}$~\cite{BES},
as this has purportedly been obtained subtracting the $\rho$ contribution to the form factor.
Our analysis above demonstrates that such an attempt does not make sense: there is no theoretically
sound way to separate the $\rho$ resonance from the nonresonant $\pi\pi$ background, 
and we have demonstrated that the $\pi\pi$ contribution to the form factor normalization is actually
dominant.  In particular also the energy region below the $\rho$ mass would have to be heavily affected 
by such a subtraction, leading to a form factor normalization that is in stark contradiction with the
$\Jp\gamma$ decay rate.  This is obviously quantitatively different from removing 
the isoscalar $\omega$ and $\phi$ resonances
from $J/\psi \to \eta,\,\eta' \gamma^*$ transition form factors due to the overall smallness of their contribution.

It would be interesting and most desirable to experimentally extract the full, unchanged, transition form factor
without any parts subtracted,
given that it is precisely the interplay between three energy regions of 
the $\Jp\gamma^*$ form factor that is most challenging theoretically: low energies below $1\GeV$ 
with the dominance of the $\rho$; potentially sizable contributions of excited light $\rho'$ resonances between
$1$ and $2\GeV$; and the contribution from charmonium in the spectral function most visible near the upper 
limit of the decay region.

%\vfill\eject

\section{Summary}
To summarize, we have analyzed the $\Jp \gamma^*$ transition form factor using dispersion theory.
We have shown that the corresponding spectral function is dominated by the $\pi^+\pi^-$ intermediate
state, of which the contribution can be calculated using the $\J\to3\pi$ $P$-wave decay amplitude as well
as the pion vector form factor.  A sum rule for the form factor normalization,
which determines the decay rate $\Jp\gamma$, is saturated to about 80\% by the $\pi\pi$ contribution only,
showing that this transition form factor is dominated by light-quark dynamics.
We have given rough estimates for the contributions of four pions, approximated by an effective 
$\rho'(1450)$ resonance, and charmonium states, comparing to the latter's (dominant) effect 
on the $J/\psi \to \eta,\,\eta' \gamma^*$ transitions.

For the differential decay rates $\Jp \ell^+\ell^-$, the $\rho$ resonance in the $\pi\pi$ spectrum
is the dominating feature, leading to very stable values for the integrated branching fractions
despite large form factor uncertainties at high energies.  
An experimental confirmation of the decay spectra predicted here,
as well as a determination of the branching fractions taking the full, unmodified form factor into account,
would be highly desirable.

\begin{acknowledgments}
We would like to thank Yun-Hua Chen, Feng-Kun Guo, and Christoph Hanhart for valuable discussions, 
and Yun-Hua Chen for useful comments on the manuscript.
Financial support by DFG and the NSFC through funds provided to the Sino-German 
CRC~110 
``Symmetries and the Emergence of Structure in QCD,''
as well as by the project ``Study of Strongly Interacting Matter'' 
(HadronPhysics3, Grant Agreement No.~283286) 
under the 7th Framework Program of the EU
is gratefully acknowledged.
\end{acknowledgments}

\end{document}